\documentclass[12pt]{article}
\usepackage{graphicx}
\usepackage{url} % not crucial - just used below for the URL 
\usepackage{caption}
\usepackage[round]{natbib}
\usepackage{amsmath,amssymb,amsfonts} % Typical maths resource packages
\usepackage{amsthm}                 % Packages to allow inclusion of graphics
\usepackage{color}                    % For creating coloured text and background
\usepackage{bm}
\usepackage{titletoc}
\usepackage{minitoc,lipsum}
\usepackage{xr}
\usepackage{textcomp, latexsym, parskip,subfig,pdflscape,fancyhdr}
\usepackage{setspace}
\usepackage{threeparttable}
\usepackage{lineno,hyperref}
\usepackage{graphicx}
\usepackage{enumerate}
\usepackage{bm}
\usepackage{mathrsfs}
\usepackage{amsmath}
\usepackage{amsfonts,amssymb}
\usepackage{diagbox}
\usepackage{booktabs}
\usepackage{multirow}
\usepackage{threeparttable}
\usepackage[svgnames]{xcolor}
\usepackage{listings}
\usepackage{tikz}
\usepackage{indentfirst}
\usepackage{expl3}
\usepackage{float}
\usepackage[title]{appendix}

\newcommand{\blind}{0}

\def\bx{ {\bm x} }

\begin{document}

	\date{}
	\def\spacingset#1{\renewcommand{\baselinestretch}%
		{#1}\small\normalsize} \spacingset{1}

	%%%%%%%%%%%%%%%%%%%%%%%%%%%%%%%%%%%%%%%%%%%%%%%%%%%%%%%%%%%%%%%%%%%%%%%%%%%%%%

	\if0\blind
	{
		\title{\bf Minimum Area Confidence Set Optimality for Simultaneous Confidence Bands for Percentiles in Linear Regression}
\author{
	Lingjiao Wang$^{1}$, Yang Han$^{1}$, Wei Liu$^{2}$,
	Frank Bretz$^{3}$\\
	$^{1}$Department of Mathematics, University of Manchester, UK \\
	$^{2}$Southampton Statistical Sciences Research Institute and School of Mathematics,\\
	University of Southampton, UK\\
	%	yang.han@manchester.ac.uk\\
	%w.liu@maths.soton.ac.uk\\
	%frank.bretz@novartis.com%\\	
	$^{3}$Novartis Pharma AG,	Basel, 4002, Switzerland
}
		\maketitle
	} \fi
	
	\if1\blind
	{
		\bigskip
		\bigskip
		\bigskip
		\begin{center}
			{\LARGE\bf Minimum Volume Confidence Set Optimality for Simultaneous Confidence Bands for Percentiles in Linear Regression}
		\end{center}
		\medskip
	} \fi

	\bigskip
	\begin{abstract}
%		Simultaneous confidence bands (SCBs) for percentiles in linear regression are useful tools with many applications. 
%		%, and the average bandwidth is widely used as a criterion for comparing different confidence bands.  
%		In this paper, the area of the confidence set for the pivotal quantities, which are generated from the confidence set of unknown parameters via a linear transformation, is utilsed as a new criterion for comparing SCBs for percentiles, named minimum area confidence set (MACS)  criterion. 
%		The MACS criterion is then used for the construction exact SCBs over any finite covariate intervals and for the comparison of several SCBs of different forms, and the optimal SCBs are determined. 
%		
%		It is discovered that the area of confidence sets for the pivotal quantities of an asymmetric SCB is uniformly and can be very substantially smaller than that of the corresponding  symmetric SCB. Therefore, under the MACS criterion, asymmetric SCBs should always be used. Furthermore, a new computationally efficient method is given to calculate the critical constants of SCBs. A real data example on drug stability study is included for illustration. 
%		
Simultaneous confidence bands (SCBs) for percentiles in linear regression are valuable tools with many applications. In this paper, we propose a novel criterion for comparing SCBs for percentiles, termed the Minimum Area Confidence Set (MACS) criterion. This criterion utilizes the area of the confidence set for the pivotal quantities, which are generated from the confidence set of the unknown parameters. Subsequently, we employ the MACS criterion to construct exact SCBs over any finite covariate intervals and to compare multiple SCBs of different forms. This approach can be used to determine the optimal SCBs.

It is discovered that the area of the confidence set for the pivotal quantities of an asymmetric SCB is uniformly and can be very substantially smaller than that of the corresponding symmetric SCB. Therefore, under the MACS criterion, exact asymmetric SCBs should always be preferred. Furthermore, a new computationally efficient method is proposed to calculate the critical constants of exact SCBs for percentiles. A real data example on drug stability study is provided for illustration.

	\end{abstract}
	
	\noindent%
	\vfill
	
	\newpage
	\spacingset{1.8} % DON'T change the spacing!
\section{Introduction}

In this paper, we consider the simple linear regression model with mean-centred covariates,
\begin{equation*}
	y_i=\bx_i^T\bm{\beta}+\epsilon_i=\beta_0+\beta_1(x_i-\bar{x})+\epsilon_i,\ i=1,2,\dots,n,
\end{equation*}
where $\bar{x}=\frac{1}{n}\sum_{i=1}^{n}x_i$, $\bx_i=(1,x_i-\bar{x})^T$, $\bm{\beta}=(\beta_0,\beta_1)^T$, the random errors $\epsilon_i$ are identically and independently distributed as $N(0,\sigma^2)$, and $\beta_0$, $\beta_1$ and $\sigma^2$ are unknown parameters. Let $\bm{X}$ denote the centered design matrix, the $i$th row of which is given by $\bm{x}_i^T$, $i=1,2,\dots,n$. 
Let $S_{xx}=\sum_{i=1}^{n}(x_i-\bar{x})^2$, then $(\bm{X}^T\bm{X})^{-1}$ is a $2\times2$ diagonal matrix given by
\begin{equation}
	(\bm{X}^T\bm{X})^{-1}=\begin{pmatrix} 1/n & 0 \\ 0 & 1/S_{xx} \end{pmatrix}.
	\label{variance matrix}
\end{equation}
Denote the least square estimators of $\bm{\beta}$ and $\sigma^2$ by $\hat{\bm{\beta}}=(\hat{\beta}_0,\hat{\beta}_1)^T=(\bm{X}^T\bm{X})^{-1}\bm{X}\bm{y}$ and $\hat{\sigma}^2=\frac{(\bm{y}-\hat{\bm{y}})^T(\bm{y}-\hat{\bm{y}})}{\nu}$ with $\nu=n-2$, $\bm{y}=(y_1,\dots,y_n)^T$ and $\hat{\bm{y}}=\bx^T\hat{\bm{\beta}}$, respectively. Then, $\hat{\bm{\beta}}\sim N_2(\bm{\beta},\sigma^2(\bm{X}^T\bm{X})^{-1})$, $\hat{\sigma}/\sigma\sim \sqrt{\chi^2_{\nu}/\nu}$, and $\hat{\bm{\beta}}$ and $\hat{\sigma}$ are independent random variables.

\subsection{Simultaneous Confidence Band for Percentiles in Linear Regression}
For linear regression model, the general $100\gamma$th percentile regression line is
\begin{equation*}
	\bm{x}^T\bm{\beta}+z_{\gamma}\sigma,
\end{equation*}
where $z_{\gamma}$ is the $100\gamma$th percentile of the standard normal distribution, i.e., $\Phi(z_{\gamma})=\int_{-\infty}^{z_{\gamma}}\phi(x)dx=\gamma$ with $\phi(x)=\exp(-x^2/2)/\sqrt{2\pi}$. 
It is noteworthy that the mean regression line $\bm{x}^T\bm{\beta}$ is the 50th percentile, which is a special case of the percentile line $\bm{x}^T\bm{\beta}+z_{\gamma}\sigma$ with $\gamma=0.5$. 
Several articles, including Spurrier (1999), Al-Saidy {\it et al.} (2003), Liu {\it et al.} (2004, 2007, 2009) and Piegorsch {\it et al.} (2005), have studied simultaneous confidence bands (SCBs) for $\bm{x}^T\bm{\beta}$. 
SCBs for the $100\gamma$th percentile line $\bm{x}^T\bm{\beta}+z_{\gamma}\sigma$ over the whole covariate range $(-\infty,\infty)$, have been studied by Steinhorst and Bowden (1971), Turner and Bowden (1977, 1979) and Thomas and Thomas (1986). 
It is known that, linear regression models are often constructed for a finite covariate range, and an exact SCB over a finite interval $(a,b)$ can be substantially narrower than a conservative SCB which is constructed over the entire covariate range $(-\infty,\infty)$ in terms of average band width, and so exact SCBs are more informative. 
Han {\it et al.} (2015) has proposed a method for constructing exact asymmetric SCBs for percentiles over a finite interval $(a,b)$.

In this paper, we focus on the SCBs $(l(x),u(x))$ for the $100\gamma$th percentile line $\bm{x}^T\bm{\beta}+z_{\gamma}\sigma$, over a finite interval of interest $x-\bar{x}\in(a,b)$, that have confidence level equal to $1-\alpha$:
\begin{equation}
	\inf_{-\infty<\beta_0,\beta_1<\infty,\ \sigma>0}P\Bigl\{l(x)\leq \bm{x}^T\bm{\beta}+z_{\gamma}\sigma \leq u(x)\ \text{for all}\ x-\bar{x}\in(a,b)\Bigr\}=1-\alpha.
	\label{condition}
\end{equation}

In drug stability studies, the $100\gamma$th percentile line can be more important than the mean regression line $\bm{x}^T\bm{\beta}$. 
In order to measure the degradation of active pharmaceutical ingredients over time, drug stability studies are commonly conducted in the pharmaceutical industry. 
For example, it is expected that drug content of a large proportion of dosage units, say $95\%$ of tablets, should be larger than the threshold $h$ (in percentage) before an expected expiry date. 
Hence, the percentile line with $\gamma=1-95\%=5\%$ is of interest. 
In this case, a two-sided SCB with a specific confidence level (e.g., $1-\alpha=95\%$) for the $5\%$-percentile line should be used to determine the expiry date. 
%For the given time interval $(a,b)$, we could compare the lower part of this two-sided SCB with threshold $h$ to assess whether no more than $5\%$ of the dosage units have the drug content below $h$ for any given point in $(a,b)$. 
%Similarly, we can also use the upper part of the SCB to get relevant information.

Consider two-sided SCBs of the form
\begin{align}
	\bm{x}^T\hat{\bm{\beta}}+z_{\gamma}\hat{\sigma}/\theta-&c_1\hat{\sigma}\sqrt{\bm{x}^T(\bm{X}^T\bm{X})^{-1}\bm{x}+(z_{\gamma})^2\xi}\leq\bm{x}^T\bm{\beta}+z_{\gamma}\sigma\nonumber\\
	&\leq \bm{x}^T\hat{\bm{\beta}}+z_{\gamma}\hat{\sigma}/\theta+c_2\hat{\sigma}\sqrt{\bm{x}^T(\bm{X}^T\bm{X})^{-1}\bm{x}+(z_{\gamma})^2\xi}\qquad\text{for all}\ x-\bar{x}\in (a,b),
	\label{band form}
\end{align}
where $c_1$ and $c_2$ denote the critical constants satisfy $(\ref{condition})$, and the constants $\theta\neq0$ and $\xi$ are selected for constructing different forms of SCBs. 
The symmetric SCB, with $c_1=c_2=c$, is a special case of SCBs in \eqref{band form}. 
In this paper, we consider six different forms of SCBs, which are the only forms available in the literature; see Steinhorst and Bowden (1971), Turner and Bowden (1977), Thomas and Thomas (1986), and Han et al. (2015). 
Table $\ref{type1}$ gives the values of $\xi$ and $\theta$ for these six bands: SB, TBU and TBE (Type $I$ with $\xi=0$), V, UV and TT (Type $II$ with $\xi\neq0$). 
The corresponding asymmetric bands are denoted as SBa, TBUa, TBEa, Va, UVa and TTa, respectively.

\begin{table}[h]
	\centering
	\caption{\label{type1}Six simultaneous confidence bands}
	\begin{threeparttable}
		\setlength{\tabcolsep}{4mm}{
			\begin{tabular}{ccccc}
				\hline
				{Name}&{Type}&{$\xi$}&{$\theta$}&{Origin}\\
				\hline
				SB&$\uppercase\expandafter{\romannumeral1}$&0&1&Steinhorst and Bowden (1971)\\
				TBU&$\uppercase\expandafter{\romannumeral1}$&0&$\sqrt{\frac{2}{v}}\frac{\Gamma(\frac{v+1}{2})}{\Gamma(\frac{v}{2})}$&Turner and Bowden (1977)\\
				TBE&$\uppercase\expandafter{\romannumeral1}$&0&$\sqrt{\frac{2}{v}}\frac{\Gamma(\frac{v}{2})}{\Gamma(\frac{v-1}{2})}$&Turner and Bowden (1977)\\
				V&$\uppercase\expandafter{\romannumeral2}$&$1-\frac{2}{v}\Big(\frac{\Gamma(\frac{v+1}{2})}{\Gamma(\frac{v}{2})}\Big)^2$&1&Han et al. (2015)\\
				UV&$\uppercase\expandafter{\romannumeral2}$&$\frac{v}{2}\Big(\frac{\Gamma(\frac{v+1}{2})}{\Gamma(\frac{v}{2})}\Big)^2-1$&$\sqrt{\frac{2}{v}}\frac{\Gamma(\frac{v+1}{2})}{\Gamma(\frac{v}{2})}$&Han et al. (2015)\\
				TT&$\uppercase\expandafter{\romannumeral2}$&$\frac{1}{2v}$&$\frac{4v-1}{4v}$&Thomas and Thomas (1986)\\
				\hline
		\end{tabular}}
	\end{threeparttable}
\end{table}

The pursuit of an optimal SCB is motivated by the desire to identify the most informative SCB among the different available forms given above.
The average width (AW) of a SCB has been widely used as an optimality criterion for comparing different forms of confidence bands since Gafarian (1964). 
In the context of percentiles of linear regression, Han et al. (2015) has conducted a thorough comparison of SCBs under the AW criterion. 
However, one drawback of the AW criterion, as pointed out in Liu and Hayter (2007),  is that it may give too much weight to the interval of interest $(a,b)$ on which the confidence band is presented. 
Hence, we consider the comparison under the minimum area confidence set (MACS) criterion.
%MACS criterion, which is proposed by Liu and Hyter (2007), overcomes the drawback of the AW criterion. 
Several authors have studied optimal confidence bands for the mean regression line, $\bm{x}^T\bm{\beta}$, using the MACS criterion based on the confidence set of $\bm{\beta}$ only; see Liu and Hayter (2007), Liu {\it et al.} (2008), and Liu and Ah-kine (2010).
Liu and Hayter (2007) has employed the MACS criterion to compare simultaneous confidence bands (SCBs) in the context of simple linear regression, and identified the optimal SCB for various scenarios. 
Additionally, Liu and Ah-kine (2010) focuses on finding the best inner-hyperbolic band for the simple linear regression model utilising the MACS criterion.
In this paper, we construct pivotal quantities that are generated from the unknown parameters $(\bm{\beta},\sigma)$ via a linear transformation. 
The area of the confidence set of the pivotal quantities is used as the new MACS criterion, and we aim to find the optimal SCBs that minimise the area of corresponding confidence set. 
Since the comparison of SCBs for percentiles in the linear regression has not been conducted under the MACS criterion, this paper aims to fill this gap. 
We also propose an efficient method to calculate critical constants of SCBs for percentiles, and this method can significantly reduce computational costs.

The layout of the paper is as follows. 
Section \ref{CS for SCB} considers the construction of confidence sets for several SCBs for percentile line, including Type $I$ and Type $II$ bands. 
In Section 3, the comparison of different forms of SCBs,  symmetric and asymmetric, for percentile lines is conducted under MACS criterion. 
In Section 4, an illustrative example on the application of SCBs for percentile lines is provided. 
Finally, Section 5 contains conclusions and discussions.

\section{Confidence Sets for Two-sided SCBs}
\label{CS for SCB}

Let $\bm{P}$ be the square root matrix of $(\bm{X}^T\bm{X})^{-1}$ given by 
\begin{equation*}
	\bm{P}=(\bm{X}^T\bm{X})^{-1/2}=\begin{pmatrix} \sqrt{1/n} & 0 \\ 0 & \sqrt{1/S_{xx}}, \end{pmatrix}
\end{equation*}
and $\bm{P}\bm{x}=(\frac{1}{\sqrt{n}},\frac{x-\bar{x}}{\sqrt{S_{xx}}})^T$. 
From \eqref{band form}, the confidence level of a two-sided SCB is given by
\begin{align}
	1-\alpha&=P\Bigl\{\bm{x}^T\hat{\bm{\beta}}+z_{\gamma}\hat{\sigma}/\theta- c_1\hat{\sigma}\sqrt{\bm{x}^T(\bm{X}^T\bm{X})^{-1}\bm{x}+(z_{\gamma})^2\xi}\leq\bm{x}^T\bm{\beta}+z_{\gamma}\sigma\nonumber\\&\qquad\leq\bm{x}^T\hat{\bm{\beta}}+z_{\gamma}\hat{\sigma}/\theta+ c_2\hat{\sigma}\sqrt{\bm{x}^T(\bm{X}^T\bm{X})^{-1}\bm{x}+(z_{\gamma})^2\xi}\quad\text{for all}\ x-\bar{x}\in (a,b)\Bigr\}\nonumber\\
	&=P\Bigl\{-c_2\leq\frac{(\bm{P}\bm{x})^T\bm{N}/U+z_{\gamma}(1/\theta-1/U)}{\sqrt{(\bm{P}\bm{x})^T(\bm{P}\bm{x})+(z_{\gamma})^2\xi}}\leq c_1\quad\text{for all}\ x-\bar{x}\in (a,b)\Bigr\}\nonumber\\
	&=P\Bigl\{-c_2\leq\frac{(\frac{N_1}{\sqrt{n}}+\frac{N_2(x-\bar{x})}{\sqrt{S_{xx}}})/U+z_{\gamma}(1/\theta-1/U)}{\sqrt{\frac{1}{n}+\frac{(x-\bar{x})^2}{S_{xx}}+z_{\gamma}^2\xi}}\leq c_1\quad\text{for all}\ x-\bar{x}\in (a,b)\Bigr\}\nonumber\\
	&=P\Biggl\{-c_2\leq \frac{\bm{w}^T(\bm{T}+\bm{\nu})}{\left \|\bm{w}  \right\|}\leq c_1\quad\text{for all}\ x-\bar{x}\in (a,b)\Biggr\}\nonumber\\
	&=P\Biggl\{-c_2\leq \frac{\bm{w}^T\bm{V}}{\left \|\bm{w}  \right\|}\leq c_1\quad\text{for all}\ x-\bar{x}\in (a,b)\Biggr\},
	\label{mixed asymmetric}
\end{align}
where $\bm{w}=\left(\sqrt{\frac{1}{n}+z_{\gamma}^2\xi},\frac{x-\bar{x}}{\sqrt{S_{xx}}}\right)^T$, $\bm{N}:=(N_1,N_2)^T=\bm{P}^{-1}(\hat{\bm{\beta}}-\bm{\beta})/\sigma\sim\bm{N}_2(\bm{0},\bm{I}_2)$, $U:=\hat{\sigma}/\sigma\sim\sqrt{\chi_{\upsilon}^2/\upsilon}$, and two vectors of pivotal quantities 
\begin{align}
	\bm{T}&=\begin{pmatrix} t_{1}\\ t_{2} \end{pmatrix}=\begin{pmatrix} \frac{N_1-\sqrt{n}z_{\gamma}}{U\sqrt{1+nz_{\gamma}^2\xi}} \\ N_2/U  \end{pmatrix},\  \bm{\nu}=\begin{pmatrix} \frac{z_{\gamma}\sqrt{n}}{\sqrt{\theta^2(1+nz_{\gamma}^2\xi)}}\\ 0  \end{pmatrix}\nonumber\\
	\text{and}\quad \bm{V}&=\begin{pmatrix} v_{1}\\ v_{2}  \end{pmatrix}=\begin{pmatrix} \frac{N_1-\sqrt{n}z_{\gamma}}{U\sqrt{1+nz_{\gamma}^2\xi}}+\frac{z_{\gamma}\sqrt{n}}{\sqrt{\theta^2(1+nz_{\gamma}^2\xi)}} \\ N_2/U  \end{pmatrix}.
	\label{sufficient statistic 2}
\end{align}

\subsection{MACS Criterion}

The MACS criterion is based on the area of the confidence set for the unknown parameter vector $\bm{\theta}=(\bm{\beta}^T,\sigma)^T=(\beta_0,\beta_1,\sigma)^T$. 
Intuitively, each $\bm{\theta}$ in a confidence set corresponds to the $\gamma$th percentile line $\bx^T\bm{\beta}+z_{\gamma}\sigma$ in linear regression that lies completely inside the corresponding confidence band and vice versa. 
Since each line $\bx^T\bm{\beta}+z_{\gamma}\sigma$ lying completely inside a confidence band is deemed by the band to be a plausible candidate for the true but unknown regression line, the smaller the area of the confidence set, the better the corresponding confidence band. 

Denote $\bm{R}_{\bm{V}}$, a region  of $\bm{V}$, by
\begin{equation}
	\bm{R}_{\bm{V}}=\Bigl\{\bm{V}: -c_2\leq\frac{\bm{w}^T\bm{V}}{\left\|\bm{w}\right\|}\leq c_1\quad\text{for all}\ x-\bar{x}\in (a,b)\Bigr\}\subset \mathcal{R}^2.
	\label{Rc asymmetric}
\end{equation}
Then, the corresponding confidence set for pivotal quantity $\bm{V}$ is
\begin{equation*}
	C(\bm{V})=\Biggl\{\bm{V}: \bm{V}\in\bm{R}_{\bm{V}}\Biggr\}.
\end{equation*}
Further, the confidence set for the pivotal quantity $\bm{T}$ is given by 
\begin{align*}
	C(\bm{T})&=\Biggl\{\bm{T}: \bm{T}+\bm{\nu}\in\bm{R}_{\bm{V}}\Biggr\}\nonumber\\
	&=\Biggl\{\bm{T}: \bm{T}\in\bm{R}_{\bm{V}}-\bm{\nu}\Biggr\},
\end{align*}
and so the corresponding confidence set for unknown parameters $\bm{\theta}$ is
\begin{align}
	C_{\bm{\theta}}&=\left\{\bm{\theta}: \bm{M}\bm{U}(\bm{a}-\bm{\theta})+\bm{\nu} \in\bm{R}_{\bm{V}}\right\}\nonumber\\
	&=\left\{\bm{\theta}: \bm{M}\bm{U}(\bm{a}-\bm{\theta}) \in\bm{R}_{\bm{V}}-\bm{\nu}\right\}
	\label{cs for theta}
\end{align}
where $\bm{a}=(\hat{\beta}_0,\hat{\beta}_1,0)^T$, 
\begin{equation*}
	\bm{M}=\begin{pmatrix} \sqrt{\frac{1}{1+nz_{\gamma}^2\xi}} & 0\\ 0 & 1  \end{pmatrix}\quad\text{and}\quad\bm{U}=\frac{1}{\hat{\sigma}}\begin{pmatrix} \sqrt{n} & 0 & z_{\gamma}\sqrt{n}\\0& \sqrt{S_{xx}}&0 \end{pmatrix}.
\end{equation*}
Let  $\bm{T}^*=\bm{U}(\bm{a}-\bm{\theta})$, then the corresponding confidence set of $\bm{T}^*$ is 
\begin{align*}
	C(\bm{T}^*)&=\Biggl\{\bm{T}^*: \bm{M}\bm{T}^*+\bm{\nu}\in\bm{R}_{\bm{V}}\Biggr\}\nonumber\\
	&=\Biggl\{\bm{T}^*: \bm{T}^*\in\bm{M}^{-1}\bm{R}_{\bm{V}}-\bm{M}^{-1}\bm{\nu}\Biggr\},
\end{align*}
and so
\begin{equation}
	\text{Area}(C(\bm{T}^*))=|\bm{M}^{-1}|\text{Area}(\bm{R}_{\bm{V}})=\text{Area}(\bm{R}_{\bm{V}})\sqrt{1+nz_{\gamma}^2\xi},
	\label{area of confidence set}
\end{equation}
where $\text{Area}(\cdot)$ is the area of a confidence set.
It is noteworthy that the confidence set $C(\bm{T}^*)$ is generated from the confidence set $C_{\bm{\theta}}$ via a linear transformation $\bm{U}(\bm{a}-\bm{\theta})$ regardless the choices of $(\xi,\theta)$. 
%Hence, the MACS criterion in this paper is based on the areas of $C(\bm{T}^*)$ for statistic $\bm{T}^*$ through the areas of $\bm{R}_{\bm{V}}$ for statistic $\bm{V}$. 

The optimisation of SCBs by MACS criterion is that of finding the region $\bm{R}_{\bm{V}}$ that minimises the $\text{Area}(C(\bm{T}^*))$ among all the regions in $\mathcal{R}^2$, such that the probability of the pivotal quantity $\bm{V}$ in $\bm{R}_{\bm{V}}$ is equal to $1-\alpha$. 
That is to say
\begin{equation}
	\min_{C(\bm{T}^*)}\int_{C(\bm{T}^*)}1d\bm{t}^*\quad\text{subject to}\quad P(\bm{V}\in\bm{R}_{\bm{V}})=\int_{\bm{R}_{\bm{V}}}f(\bm{v})d\bm{v}=1-\alpha,
	\label{minimum volum1}
\end{equation}
where $f(\bm{v})$ is the probability density function (pdf) of  $\bm{V}$, see Appendix. 

\begin{figure}[h]
	\centering
	\includegraphics[width=10cm,height=9cm]{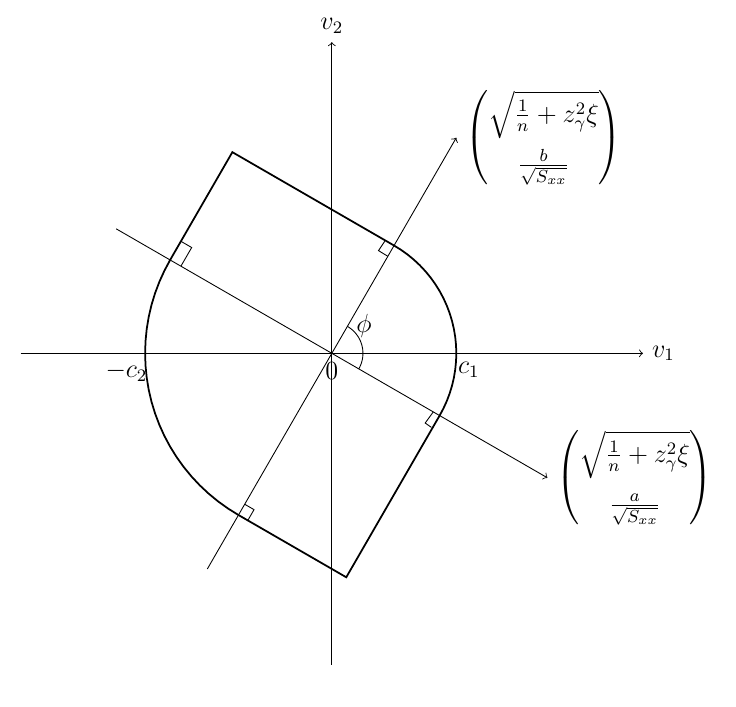}
	\caption{The area of $\bm{R}_{\bm{V}}$}
	\label{confidence set2}
\end{figure}

From \eqref{Rc asymmetric}, $\bm{R}_{\bm{V}}$ is spindle-shaped as illustrated in Figure $\ref{confidence set2}$, where the angle $\phi$ is formed by the vectors $\begin{pmatrix} \sqrt{\frac{1}{n}+z_{\gamma}^2\xi}\\ \frac{a}{\sqrt{S_{xx}}}  \end{pmatrix}$ and $\begin{pmatrix} \sqrt{\frac{1}{n}+z_{\gamma}^2\xi}\\ \frac{b}{\sqrt{S_{xx}}}  \end{pmatrix}$, and can be calculated by
\begin{align}
	\cos(\phi)&=\begin{pmatrix} \sqrt{\frac{1}{n}+z_{\gamma}^2\xi}\\ \frac{a}{\sqrt{S_{xx}}}  \end{pmatrix}^T\begin{pmatrix} \sqrt{\frac{1}{n}+z_{\gamma}^2\xi}\\ \frac{b}{\sqrt{S_{xx}}}  \end{pmatrix}\Bigg/\left \|\begin{pmatrix} \sqrt{\frac{1}{n}+z_{\gamma}^2\xi}\\ \frac{a}{\sqrt{S_{xx}}}  \end{pmatrix}  \right\|\left \|\begin{pmatrix} \sqrt{\frac{1}{n}+z_{\gamma}^2\xi}\\ \frac{b}{\sqrt{S_{xx}}}  \end{pmatrix}  \right\|\nonumber\\
	&=\frac{\frac{1}{n}+z_{\gamma}^2\xi+\frac{ab}{S_{xx}}}{\sqrt{\Big[\frac{1}{n}+z_{\gamma}^2\xi+\frac{a^2}{S_{xx}}\Big]\Big[\frac{1}{n}+z_{\gamma}^2\xi+\frac{b^2}{S_{xx}}\Big]}}.
	\label{cosine phi1}
\end{align}
When $(b-a)$ increases, $\phi$ increases.

Set $c_{min}=\min(c_1,c_2)$ and $c_{max}=\max(c_1,c_2)$. The $\text{Area}(\bm{R}_{\bm{V}})$ in \eqref{area of confidence set} can be calculated for the following situations 
\begin{enumerate}
	\item[1.]  When $\phi<\pi/2$,
	$\text{Area}(\bm{R}_{\bm{V}})=(c_{min}/\cos\phi+c_{max})^2/\tan\phi-c_{min}^2\tan\phi+\phi(c_1^2+c_2^2)/2.$
	\item[2.] When $\phi=\pi/2$, $\text{Area}(\bm{R}_{\bm{V}})=2c_1c_2+\phi(c_1^2+c_2^2)/2.$
	\item[3.] When $\phi>\pi/2$, 
	\begin{enumerate}
		\item[(1)] if $c_{min}/\cos(\pi-\phi)>c_{max}$, $\text{Area}(\bm{R}_{\bm{V}})=c_{min}^2\tan(\pi-\phi)-(c_{min}/\cos(\pi-\phi)-c_{max})^2/\tan(\pi-\phi)+\phi(c_1^2+c_2^2)/2$;
		\item[(2)] if $c_{min}/\cos(\pi-\phi)\leq c_{max}$, $\text{Area}(\bm{R}_{\bm{V}})=c_{min}\sqrt{c_{max}^2-c_{min}^2}+c_{min}^2\phi/2+c_{max}^2(2\pi-\phi-2\arccos(c_{min}/c_{max}))/2$.
	\end{enumerate}
\end{enumerate}

\subsection{Confidence Level of Confidence Set}

From the construction of SCBs, the critical constants $c_1$ and $c_2$ are determined from $P\{\bm{V}\in\bm{R_V}\}=1-\alpha$. 
The confidence set $\bm{R_V}$ in Figure \ref{confidence set2} is partitioned into four triangles $\bm{R}_{\bm{V};M_{i}}$ and four fans $\bm{R}_{\bm{V};N_{i}}$ ($i=1,2,3,4$), see Figure 2. 
Here, $\phi$ in \eqref{cosine phi1} is partitioned into $\phi_1$ and $\phi_2$. 
The angles $\phi_1$ and $\phi_2$ can be calculated by 
\begin{align*}
	\cos(\phi_1)&=\begin{pmatrix} \sqrt{\frac{1}{n}+z_{\gamma}^2\xi}\\ 0  \end{pmatrix}^T\begin{pmatrix} \sqrt{\frac{1}{n}+z_{\gamma}^2\xi}\\ \frac{b}{\sqrt{S_{xx}}}  \end{pmatrix}\Bigg/\left \|\begin{pmatrix} \sqrt{\frac{1}{n}+z_{\gamma}^2\xi}\\ 0  \end{pmatrix}  \right\|\left \|\begin{pmatrix} \sqrt{\frac{1}{n}+z_{\gamma}^2\xi}\\ \frac{b}{\sqrt{S_{xx}}}  \end{pmatrix}  \right\|\nonumber\\
	&=\frac{\frac{1}{n}+z_{\gamma}^2\xi}{\sqrt{[\frac{1}{n}+z_{\gamma}^2\xi]\Big[\frac{1}{n}+z_{\gamma}^2\xi+\frac{b^2}{S_{xx}}\Big]}},\nonumber\\
	\text{and}\ \cos(\phi_2)&=\begin{pmatrix} \sqrt{\frac{1}{n}+z_{\gamma}^2\xi}\\ 0  \end{pmatrix}^T\begin{pmatrix} \sqrt{\frac{1}{n}+z_{\gamma}^2\xi}\\ \frac{a}{\sqrt{S_{xx}}}  \end{pmatrix}\Bigg/\left \|\begin{pmatrix} \sqrt{\frac{1}{n}+z_{\gamma}^2\xi}\\ 0  \end{pmatrix}  \right\|\left \|\begin{pmatrix} \sqrt{\frac{1}{n}+z_{\gamma}^2\xi}\\ \frac{a}{\sqrt{S_{xx}}}  \end{pmatrix}  \right\|\nonumber\\
	&=\frac{\frac{1}{n}+z_{\gamma}^2\xi}{\sqrt{[\frac{1}{n}+z_{\gamma}^2\xi]\Big[\frac{1}{n}+z_{\gamma}^2\xi+\frac{a^2}{S_{xx}}\Big]}}.
	\label{phi_1 and phi_2}
\end{align*}
\begin{figure}[h]
	\centering
	\includegraphics[width=10cm,height=9cm]{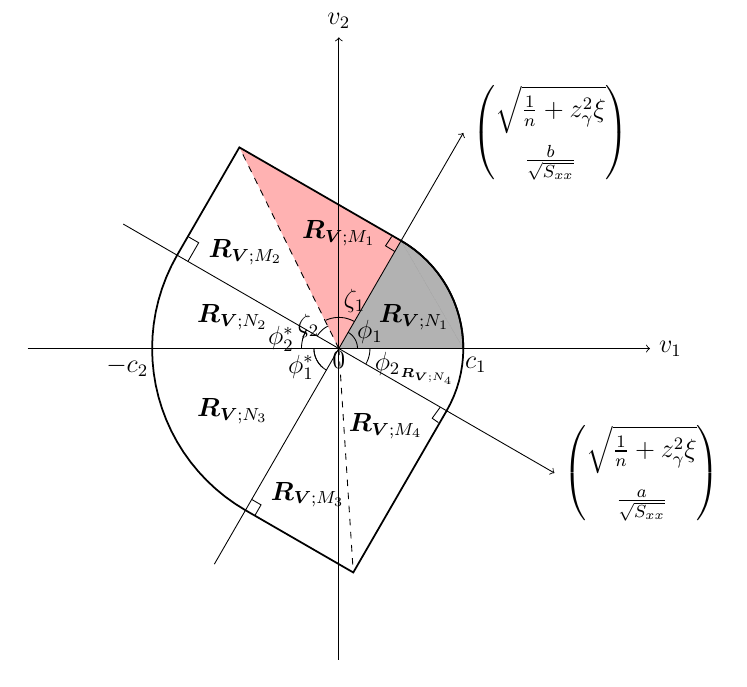}
	\caption{The partitioned regions of $\bm{R_V}$}
	\label{parts of R_v}
\end{figure}

In order to calculate the probability of $\bm{V}$ in $\bm{R_V}$, three different situations according to $\phi$ are considered below.
\begin{enumerate}
	\item[1.] When $\phi=\frac{\pi}{2}$, we have $\zeta_1=\arctan\left(\frac{c_{max}}{c_{min}}\right)$, $\zeta_2=\pi-\zeta_1-\phi$, $\phi_1^*=\phi_1$ and $\phi_2^*=\phi_2$.
	\item[2.] When $\phi<\frac{\pi}{2}$, we have $\zeta_1=\arctan\left[\left(\frac{c_{max}}{\cos\phi}+c_{min}\right)\tan\left(\frac{\pi}{2}-\phi\right)/c_{min}\right]$, $\zeta_2=\pi-\zeta_1-\phi$, $\phi_1^*=\phi_1$ and $\phi_2^*=\phi_2$.
	\item[3.] When $\phi>\frac{\pi}{2}$,
	\begin{enumerate}
		\item[(1)] if $c_{min}/\cos(\pi-\phi)>c_{max}$, we have $\zeta_1=\arctan\left[\left(\frac{c_{max}}{\cos(\pi-\phi)}-c_{min}\right)\tan\left(\phi-\frac{\pi}{2}\right)/c_{min}\right]$, $\zeta_2=\pi-\zeta_1-\phi$, $\phi_1^*=\phi_1$ and $\phi_2^*=\phi_2$;
		\item[(2)] if $c_{min}/\cos(\pi-\phi)\leq c_{max}$, we have $\zeta_1=\arccos\left(\frac{c_{min}}{c_{max}}\right)$, $\zeta_2=0$, $\phi_1^*=\pi-\phi_2-\zeta_1$, and $\phi_2^*=\pi-\phi_1-\zeta_1$.
	\end{enumerate}
\end{enumerate}

Define the polar coordinates of $\bm{V}=(V_1,V_2)^T$, $(R,\delta)$, by 
\begin{equation*}
	V_1=R\cos\delta,\ V_2=R\sin\delta\ \text{for}\ R>0\ \text{and}\ \delta\in[0,2\pi). 
\end{equation*}
The joint density of $(R,\delta)$ is
\begin{align}
	f_{R,\delta}(r,\delta)&=f_{V_1,V_2}(v_1,v_2)r\nonumber\\
	&=\frac{r\exp(-q_3^2/2)\nu^{\nu/2}}{q_1 2^{\nu/2}\pi\Gamma(\nu/2)}\int_{0}^{\infty}v_3^{\nu+1}\exp\Bigg\{-\frac{1}{2}\Bigg[\left(\left(\frac{r\cos\delta-q_2}{q_1}\right)^2+r^2(\sin\delta)^2+\nu\right)v_3^2\nonumber\\
	&\hspace{7.5cm}+\frac{2q_3(r\cos\delta-q_2)}{q_1}v_3\Bigg]\Bigg\}dv_3.
	\label{jdf for R and delta}
\end{align}
where $q_1=\sqrt{\frac{1}{1+nz_{\gamma}^2\xi}}$, $q_2=\frac{z_{\gamma}\sqrt{n}}{\sqrt{\theta^2(1+nz_{\gamma}^2\xi)}}$ and $q_3=z_{\gamma}\sqrt{n}$, the derivation of $f_{V_1,V_2}(v_1,v_2)$ in \eqref{jdf for R and delta} is given in the Appendix. 

As pointed out above region $\bm{R_V}$ is partitioned in the following way:
\begin{equation*}
	\bm{R_V}=\bm{R}_{\bm{V};M_{1}}+\bm{R}_{\bm{V};M_{2}}+\bm{R}_{\bm{V};M_{3}}+\bm{R}_{\bm{V};M_{4}}+\bm{R}_{\bm{V};N_{1}}+\bm{R}_{\bm{V};N_{2}}+\bm{R}_{\bm{V};N_{3}}+\bm{R}_{\bm{V};N_{4}}
\end{equation*}
where
\begin{align*}
	\bm{R}_{\bm{V};M_{1}}&=\{\bm{V}:\ \delta\in[\phi_1,\phi_1+\zeta_1),\ 0\leq\left(\cos\left(\phi_1\right),\sin\left(\phi_1\right)\right)\bm{V}\leq c_1\}\nonumber\\
	&=\left\{(R,\delta):\ \delta\in[\phi_1,\phi_1+\zeta_1),\ R\leq \frac{c_1}{\cos\left(\delta-\phi_1\right)}\right\},\nonumber\\
	\bm{R}_{\bm{V};M_{2}}&=\{\bm{V}:\ \delta\in[\phi_1+\zeta_1,\pi-\phi_2^*),\ 0\leq\left(\cos\left(\pi-\phi_2^*\right),\sin\left(\pi-\phi_2^*\right)\right)\bm{V}\leq c_2\}\nonumber\\
	&=\left\{(R,\delta):\ \delta\in[\phi_1+\zeta_1,\pi-\phi_2^*),\ R\leq \frac{c_2}{\cos\left(\delta-\pi+\phi_2^*\right)}\right\},\nonumber\\
	\bm{R}_{\bm{V};M_{3}}&=\{\bm{V}:\ \delta\in[\pi+\phi_1^*,\pi+\phi_1^*+\zeta_2),\ 0\leq\left(\cos\left(\pi+\phi_1^*\right),\sin\left(\pi+\phi_1^*\right)\right)\bm{V}\leq c_2\}\nonumber\\
	&=\left\{(R,\delta):\ \delta\in[\pi+\phi_1^*,\pi+\phi_1^*+\zeta_2),\ R\leq \frac{c_2}{\cos\left(\delta-\pi-\phi_1^*\right)}\right\},
	\nonumber\\
	\bm{R}_{\bm{V};M_{4}}&=\{\bm{V}:\ \delta\in[\pi+\phi_1^*+\zeta_2,2\pi-\phi_2),\ 0\leq\left(\cos\left(2\pi-\phi_2\right),\sin\left(2\pi-\phi_2\right)\right)\bm{V}\leq c_2\}\nonumber\\
	&=\left\{(R,\delta):\ \delta\in[\pi+\phi_1^*+\zeta_2,2\pi-\phi_2),\ R\leq \frac{c_2}{\cos\left(\delta-2\pi+\phi_2\right)}\right\},
	%\nonumber\\
\end{align*}
\begin{align*}
	\bm{R}_{\bm{V};N_{1}}&=\{\bm{V}:\ \delta\in[0,\phi_1),\ ||\bm{V}||\leq c_1\}=\{(R,\delta):\ \delta\in[0,\phi_1),\ R\leq c_1 \},\nonumber\\
	\bm{R}_{\bm{V};N_{2}}&=\{\bm{V}:\ \delta\in[\pi-\phi_2^*,\pi),\ ||\bm{V}||\leq c_2\}=\{(R,\delta):\ \delta\in[\pi-\phi_2^*,\pi),\ R\leq c_2 \},\nonumber\\
	\bm{R}_{\bm{V};N_{3}}&=\{\bm{V}:\ \delta\in[\pi,\pi+\phi_1^*),\ ||\bm{V}||\leq c_2\}=\{(R,\delta):\ \delta\in[\pi,\pi+\phi_1^*),\ R\leq c_2 \},\nonumber\\
	\bm{R}_{\bm{V};N_{4}}&=\{\bm{V}:\ \delta\in[2\pi-\phi_2,2\pi),\ ||\bm{V}||\leq c_2\}=\{(R,\delta):\ \delta\in[\pi,\pi+\phi_1^*),\ R\leq c_2 \}.
\end{align*}

Therefore, we have
\begin{align}
	P\{\bm{V}\in \bm{R}_{\bm{V}}\}=&P\{\bm{V}\in \bm{R}_{\bm{V};M_{1}}\}+P\{\bm{V}\in \bm{R}_{\bm{V};M_{2}}\}+P\{\bm{V}\in \bm{R}_{\bm{V};M_{3}}\}+P\{\bm{V}\in \bm{R}_{\bm{V};M_{4}}\}\nonumber\\
	&+P\{\bm{V}\in \bm{R}_{\bm{V};N_{1}}\}+P\{\bm{V}\in \bm{R}_{\bm{V};N_{2}}\}+P\{\bm{V}\in \bm{R}_{\bm{V};N_{3}}\}+P\{\bm{V}\in \bm{R}_{\bm{V};N_{4}}\}\nonumber\\
	=&\int_{\phi_1}^{\phi_1+\zeta_1}\int_{0}^{\frac{c_1}{\cos\left(\delta-\phi_1\right)}}f_{R,\delta}(r,\delta)drd\delta+\int_{\phi_1+\zeta_1}^{\pi-\phi_2^*}\int_{0}^{\frac{c_2}{\cos\left(\delta+\phi_2^*-\pi\right)}}f_{R,\delta}(r,\delta)drd\delta\nonumber\\
	&+\int_{\pi+\phi_1^*}^{\pi+\phi_1^*+\zeta_2}\int_{0}^{\frac{c_2}{\cos\left(\delta-\phi_1^*-\pi\right)}}f_{R,\delta}(r,\delta)drd\delta+\int_{\pi+\phi_1^*+\zeta_2}^{2\pi-\phi_2}\int_{0}^{\frac{c_1}{\cos\left(\delta+\phi_2-2\pi\right)}}f_{R,\delta}(r,\delta)drd\delta\nonumber\\
	&+\int_{0}^{\phi_1}\int_{0}^{c_1}f_{R,\delta}(r,\delta)drd\delta+\int_{\pi-\phi_2^*}^{\pi}\int_{0}^{c_2}f_{R,\delta}(r,\delta)drd\delta+\int_{\pi}^{\pi+\phi_1^*}\int_{0}^{c_2}f_{R,\delta}(r,\delta)drd\delta\nonumber\\
	&+\int_{2\pi-\phi_2}^{2\pi}\int_{0}^{c_1}f_{R,\delta}(r,\delta)drd\delta.
	\label{confidence level for V}
\end{align}
Expression \eqref{confidence level for V} gives the confidence level of a two-sided SCB for the percentile line with given $c_1$ and $c_2$.
%, which can be used to calculate the critical constants $c_1$ and $c_2$ for given $\alpha$ and $\gamma$.

From  \eqref{confidence level for V}, we propose a new method, based on the confidence set of $\bm{V}$, to compute the critical constants $(c_1,c_2)$ for SCBs: 
given sample size $n$, confidence level $1-\alpha$, $\gamma$, and covariate interval of interest $(a,b)$, we can compute the values of $c_1$ and $c_2$ to satisfy $P\{\bm{V}\in \bm{R}_{\bm{V}}\}= 1-\alpha$ and to minimise the area of $\bm{R_V}$ simultaneously.

From our numerical investigations, the computation of critical constants only takes 8 seconds for symmetric SCBs and 115 seconds for asymmetric SCBs on an ordinary Window’s PC (Intel(R) Core(TM) i7-6700 CPU with 3.40GHz, 3.41 GHz, RAM 16.0 GB). 
We additionally use the simulation-based method introduced by Han et al. (2015) to compute the critical constants $(c_1, c_2)$. The simulation-based method requires approximately 340 seconds  for symmetric SCBs and 500 seconds  for asymmetric SCBs, utilising 1,000,000 simulations, on the same Windows PC. 
Therefore, our newly proposed method is significantly more efficient.

\section{Comparisons under the MACS Criterion}
\label{conparison}
In our numerical comparisons, we focus on the case where $a=-b$, i.e., the interval $x-\bar{x}\in(a,b)$ is symmetric about 0. 
Let $s=b/\sqrt{S_{xx}}$. Note that, the range of $(a,b)$ is large when $s$ is large.  
In this case, the critical constants $(c_1,c_2)$ and so the areas of confidence sets for $\bm{V}$ depend only on $s$, $n$, $\gamma$, and $1-\alpha$. 
In our numerical comparison, the settings we used are as follows: (i) $\alpha=0.1,0.01$; (ii) $\gamma=0.75,0.95$; (iii) $n=10,100$; and (iv) $s=0.1,1,10$. 
According to  \eqref{cosine phi1}, the $\phi$-values for each SCB depend on $a=-b$ and $S_{xx}$ only through $s$. 
For Type $I$ bands, the $\phi$-values are denoted as $\phi^{I}$ in the Tables below. 
For Type $II$ bands, the $\phi$-values also depend on different $\xi$-values. 
Even with different $\xi$-values, the $\phi$-values for all Type $II$ bands remain the same up to two decimal places for given $s$ and $n$, which are denoted by $\phi^{II}$ in the tables below.
%Hence, we use the $\phi$-values of UV bands as the $\phi$-values for Type $II$ bands, which are denoted as $\phi^{II}$-values in the tables below. 
%In Table \ref{comparison_I}-\ref{comparison specific}, $\phi^{I}$and $\phi^{II}$ denote the $\phi$-values for Type $I$ and Type $II$ bands, respectively. 

%Here, symmetric bands UV and TBE are used as an example to illustrate the ratios of areas of confidence sets in comparison. 
Suppose we have two bands $A$ and $B$ with confidence sets $C_{A}(\bm{T}^*)$ and $C_{B}(\bm{T}^*)$, respectively. 
Given the interval of interest $(a,b)$, $n$, $\alpha$, $\gamma$, $\theta$ and $\xi$,  the ratio
\begin{equation*}
	r=\frac{\text{Area}(C_{B}(\bm{T}^*))}{\text{Area}(C_{A}(\bm{T}^*))}
	\label{r ratio MACS}
\end{equation*}
is of interest for comparing bands $A$ and $B$ under MACS criterion.
When $r>1$, the area of $C_{A}(\bm{T}^*)$ is smaller, and so $A$ is better; otherwise, $B$ is better.

%Note that the entry $r<1$ in Tables $\ref{comparison_I}$-$\ref{comparison specific}$ means a smaller confidence set and so better band. 
Table \ref{comparison_I} shows the comparison among Type $I$ bands relative to TBEa band. 
Table \ref{comparison_II} shows the comparison among Type $II$ bands relative to UVa band.
According to Table \ref{comparison_I}, asymmetric Type $I$ bands outperform symmetric Type $I$ bands by up to about 100\% when both $1-\alpha$ and $\gamma$ are large. 
Among the three symmetric Type $I$ bands, TBE has the smallest area of $C(\bm{T}^*)$, and so TBE is the best in terms of MACS. 
The difference in the areas of confidence sets among the three asymmetric Type $I$ bands is small. 
From Table \ref{comparison_II}, asymmetric Type $II$ bands are better than symmetric Type $II$ bands, by as much as about 78\% when $1-\alpha$ is large and $s$ is small. 
Among the three symmetric Type $II$ bands, V band consistently performs slightly worse than UV band under MACS criterion, and therefore, V band is not recommended. 
The differences in the areas of confidence sets between UV and TT are small, so either band can be used. 
%From hindsight, it is not surprising that UV and TT are very similar as their corresponding  values of $\xi$ and $\theta$ are very close when $n\geq 10$.
Additionally, the differences in the areas of confidence sets among the three asymmetric Type $II$ bands are minor. 
In general, an asymmetric SCB is better than the corresponding symmetric SCB according to the MACS criterion.
% and this conclusion is consistent with the comparison results in Han et al. (2015) under AW criterion.

To investigate the performances cross Type $I$ and Type $II$ bands, both symmetric and asymmetric, Table \ref{comparison specific} presents the comparison of  TBE, TBEa, UV relative to UVa. 
Notably, UVa performs the best in most cases, while TBEa is the best when the sample size is small and $x$-values in the training dataset are dispersed (i.e. $n=10$ and $s=10$). 

\begin{figure}[!h]
	\centering
	\includegraphics[width=10cm,height=6cm]{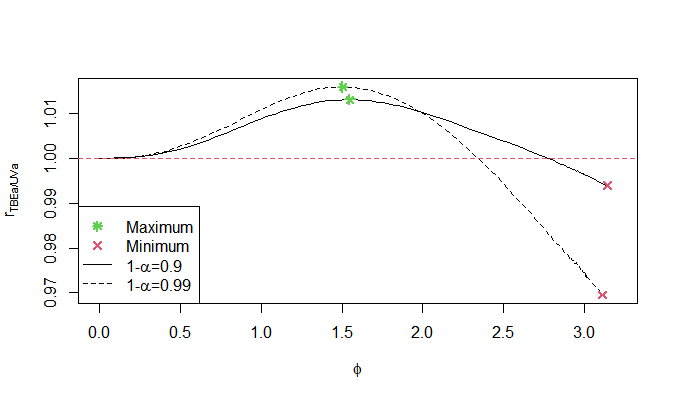}
	\caption{The ratios between TBEa and UVa bands when $\gamma=0.75$, $n=10$ and $\phi\in(0,2\pi)$}
	\label{ratios between TBEa and UVa 75}
\end{figure}

\begin{figure}[!h]
	\centering
	\includegraphics[width=10cm,height=6cm]{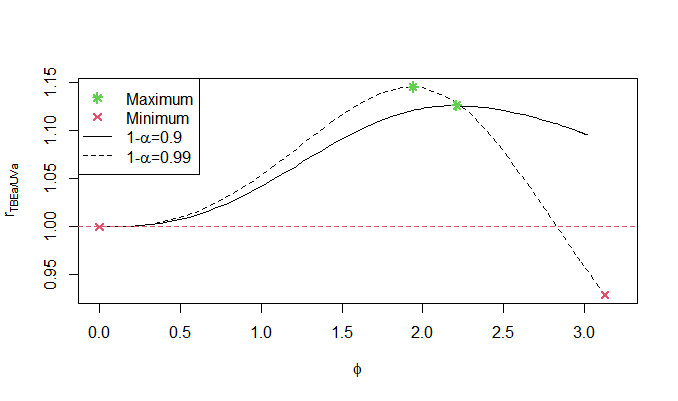}
	\caption{The ratios between TBEa and UVa bands when $\gamma=0.95$, $n=10$ and $\phi\in(0,2\pi)$}
	\label{ratios between TBEa and UVa 95}
\end{figure}

Based on the results in Table \ref{comparison specific}, we conduct a comparison between TBEa and UVa with $n=10$ in Figures \ref{ratios between TBEa and UVa 75} and \ref{ratios between TBEa and UVa 95}. 
The $\phi^{I}$-values for Type $I$ bands are utilised as $\phi$-axis. 
Figures \ref{ratios between TBEa and UVa 75} and \ref{ratios between TBEa and UVa 95} reveal that the $r$ ratio is below 1 for large $\phi$ and $1-\alpha$ (e.g., $\phi>2.8$ and $1-\alpha=0.99$), when the sample size $n$ is small.  
Therefore, one concludes that TBEa is better than UVa when both $\phi$ and $1-\alpha$ are large, and sample size $n$ is small. 
It is also noteworthy that $\phi^{I}$ and $\phi^{II}$ are larger than 3 when $s\geq 10$, indicating that $x$-values of training dataset are dispersed. 
%Hence, TBEa is better when $x$-values of training dataset are dispersed and confidence level $1-\alpha$ is large. 
Hence, asymmetric Type $I$ bands should be used only for dispersed $x$-values of training dataset and large confidence level $1-\alpha$,  when sample size $n$ is small. 
Therefore, asymmetric Type $II$ bands, like UVa, are recommanded in general.

\begin{table}[!h]
	\centering
	\caption{\label{comparison_I}Ratios $r$, relative to TBEa, of the areas of $C(\bm{T}^*)$ for symmetric and asymmetric Type $I$ SCBs: SB, TBU, TBE, SBa , TBUa.}
\begin{threeparttable}
	\begin{tabular}{cccccccccc}
		\hline
		{$1-\alpha$}&{$\gamma$}&{$n$}&{$s$}&{$\phi^{I}$}&{$SB$}&{$TBU$}&{$TBE$}&{$SBa$}&{$TBUa$}\\
		\hline
		0.9 & 0.75 & 10 & 0.1 & 0.613 & 1.038  & 1.020  & 1.000  & 1.000  & 1.000  \\ 
		~ & ~ & ~ & 1 & 2.529 & 1.039  & 1.024  & 1.003  & 1.003  & 1.001  \\ 
		~ & ~ & ~ & 10 & 3.078 & 1.038  & 1.023  & 1.002  & 1.008  & 1.005  \\
		~ & ~ & 100 & 0.1 & 1.571 & 1.004  & 1.003  & 1.002  & 1.000  & 1.000  \\ 
		~ & ~ & ~ & 1 & 2.942 & 1.002  & 1.001  & 1.000  & 1.000  & 1.000  \\
		~ & ~ & ~ & 10 & 3.122 & 1.002  & 1.001  & 1.000  & 1.001  & 1.000  \\
		~ & 0.95 & 10 & 0.1 & 0.613 & 1.158  & 1.079  & 1.000  & 0.999  & 0.999  \\ 
		~ & ~ & ~ & 1 & 2.529 & 1.158  & 1.087  & 1.004  & 0.977  & 0.985  \\ 
		~ & ~ & ~ & 10 & 3.078 & 1.160  & 1.089  & 1.006  & 0.965  & 0.974  \\ 
		~ & ~ & 100 & 0.1 & 1.571 & 1.006  & 1.003  & 1.000  & 0.999  & 1.000  \\ 
		~ & ~ & ~ & 1 & 2.942 & 1.006  & 1.003  & 1.000  & 0.996  & 0.998  \\ 
		~ & ~ & ~ & 10 & 3.122 & 1.006  & 1.003  & 1.000  & 0.996  & 0.998  \\ 
		0.99 & 0.75 & 10 & 0.1 & 0.613 & 1.322  & 1.287  & 1.206  & 1.000  & 1.000  \\ 
		~ & ~ & ~ & 1 & 2.529 & 1.243  & 1.218  & 1.160  & 1.006  & 1.004  \\ 
		~ & ~ & ~ & 10 & 3.078 & 1.226  & 1.201  & 1.144  & 1.014  & 1.009  \\
		~ & ~ & 100 & 0.1 & 1.571 & 1.025  & 1.021  & 1.014  & 1.000  & 1.000  \\
		~ & ~ & ~ & 1 & 2.942 & 1.023  & 1.019  & 1.013  & 1.000  & 1.001  \\ 
		~ & ~ & ~ & 10 & 3.122 & 1.022  & 1.019  & 1.013  & 1.001  & 1.000  \\
		~ & 0.95 & 10 & 0.1 & 0.613 & 1.807  & 1.731  & 1.554  & 1.000  & 1.000  \\ 
		~ & ~ & ~ & 1 & 2.529 & 1.921  & 1.844  & 1.664  & 0.974  & 0.980  \\ 
		~ & ~ & ~ & 10 & 3.078 & 2.004  & 1.922  & 1.735  & 1.006  & 1.000  \\ 
		~ & ~ & 100 & 0.1 & 1.571 & 1.085  & 1.071  & 1.046  & 1.000  & 1.000  \\
		~ & ~ & ~ & 1 & 2.942 & 1.107  & 1.093  & 1.067  & 0.990  & 0.993  \\ 
		~ & ~ & ~ & 10 & 3.122 & 1.114  & 1.100  & 1.074  & 0.986  & 0.989  \\ 
		\hline
	\end{tabular}
\begin{tablenotes}
	\footnotesize                     
	\item[1] $\phi^{I}$ is the angle $\phi$ in \eqref{cosine phi1} for symmetric and asymmetric Type $I$ bands ($SB$, $TBU$, $TBE$, $SBa$, $TBUa$ and $TBEa$ bands). 
\end{tablenotes}
\end{threeparttable}
\end{table}
\newpage

\begin{table}[!h]
	\centering
	\caption{\label{comparison_II}Ratios $r$, relative to UVa, of the areas of $C(\bm{T}^*)$ for symmetric and asymmetric Type $II$ SCBs: V, UV, TT, Va, TTa.}
\begin{threeparttable}
	\begin{tabular}{cccccccccc}
		\hline
		{$1-\alpha$}&{$\gamma$}&{$n$}&{$s$}&{$\phi^{II}$}&{$V$}&{$UV$}&{$TT$}&{$Va$}&{$TTa$}\\
		\hline
		0.9 & 0.75 & 10 & 0.1 & 0.543 & 1.038  & 1.020  & 1.020  & 1.000  & 1.000  \\ 
		~ & ~ & ~ & 1 & 2.451 & 1.036  & 1.022  & 1.022  & 1.002  & 1.000  \\ 
		~ & ~ & ~ & 10 & 3.070 & 1.027  & 1.014  & 1.014  & 1.004  & 1.000  \\ 
		~ & ~ & 100 & 0.1 & 1.466 & 1.002  & 1.001  & 1.001  & 1.000  & 1.000  \\ 
		~ & ~ & ~ & 1 & 2.920 & 1.002  & 1.001  & 1.001  & 1.000  & 1.000  \\ 
		~ & ~ & ~ & 10 & 3.119 & 1.001  & 1.000  & 1.000  & 1.001  & 1.000  \\ 
		~ & 0.95 & 10 & 0.1 & 0.543 & 1.161  & 1.081  & 1.080  & 1.000  & 1.000  \\ 
		~ & ~ & ~ & 1 & 2.451 & 1.141  & 1.092  & 1.091  & 1.001  & 0.999  \\ 
		~ & ~ & ~ & 10 & 3.070 & 1.114  & 1.070  & 1.068  & 1.007  & 0.998  \\ 
		~ & ~ & 100 & 0.1 & 1.466 & 1.007  & 1.003  & 1.003  & 1.000  & 1.000  \\ 
		~ & ~ & ~ & 1 & 2.920 & 1.006  & 1.003  & 1.003  & 1.001  & 1.000  \\ 
		~ & ~ & ~ & 10 & 3.119 & 1.005  & 1.002  & 1.002  & 1.001  & 1.000  \\ 
		0.99 & 0.75 & 10 & 0.1 & 0.543 & 1.322  & 1.286  & 1.286  & 1.000  & 1.000  \\ 
		~ & ~ & ~ & 1 & 2.451 & 1.191  & 1.168  & 1.169  & 1.002  & 1.000  \\ 
		~ & ~ & ~ & 10 & 3.070 & 1.149  & 1.128  & 1.128  & 1.004  & 0.999  \\ 
		~ & ~ & 100 & 0.1 & 1.466 & 1.024  & 1.021  & 1.021  & 1.000  & 1.000  \\ 
		~ & ~ & ~ & 1 & 2.920 & 1.016  & 1.014  & 1.014  & 1.001  & 1.000  \\ 
		~ & ~ & ~ & 10 & 3.119 & 1.015  & 1.012  & 1.012  & 1.000  & 1.000  \\ 
		~ & 0.95 & 10 & 0.1 & 0.543 & 1.807  & 1.731  & 1.729  & 1.000  & 1.000  \\ 
		~ & ~ & ~ & 1 & 2.451 & 1.535  & 1.476  & 1.479  & 1.003  & 0.999  \\ 
		~ & ~ & ~ & 10 & 3.070 & 1.381  & 1.328  & 1.331  & 1.006  & 0.998  \\ 
		~ & ~ & 100 & 0.1 & 1.466 & 1.085  & 1.072  & 1.072  & 1.000  & 1.000  \\ 
		~ & ~ & ~ & 1 & 2.920 & 1.059  & 1.050  & 1.050  & 1.001  & 1.000  \\ 
		~ & ~ & ~ & 10 & 3.119 & 1.055  & 1.045  & 1.045  & 1.001  & 1.000  \\  
		\hline
	\end{tabular}
\begin{tablenotes}
	\footnotesize                     
	\item[1] $\phi^{II}$ is the angle $\phi$ in \eqref{cosine phi1} for symmetric and asymmetric Type $II$ bands ($V$, $UV$, $TT$, $Va$, $UVa$ and $TTa$ bands). 
\end{tablenotes}
\end{threeparttable}
\end{table}
\newpage

\begin{table}[!h]
	\centering
	\caption{\label{comparison specific}Ratios $r$,  relative to UVa, of the areas of $C(\bm{T}^*)$ for SCBs: TBE , TBEa, UV.}
	\begin{threeparttable}
	\begin{tabular}{ccccccccc}
		\hline
		{$1-\alpha$}&{$\gamma$}&{$n$}&{$s$}&{$\phi^{I}$}&{$\phi^{II}$}&{$TBE$}&{$TBEa$}&{$UV$}\\
		\hline
		0.9 & 0.75 & 10 & 0.1 & 0.613 & 0.543 & 1.003  & 1.005  & 1.022  \\ 
		~ & ~ & ~ & 1 & 2.529 & 2.451 & 1.008  & 1.005  & 1.027  \\ 
		~ & ~ & ~ & 10 & 3.078 & 3.070 & 0.998  & 0.994  & 1.012  \\
		~ & ~ & 100 & 0.1 & 1.571 & 1.466 & 1.014  & 1.012  & 1.003  \\ 
		~ & ~ & ~ & 1 & 2.942 & 2.920 & 1.006  & 1.005  & 1.002  \\ 
		~ & ~ & ~ & 10 & 3.122 & 3.119 & 1.004  & 1.005  & 1.001  \\ 
		~ & 0.95 & 10 & 0.1 & 0.613 & 0.543 & 1.015  & 1.014  & 1.082  \\ 
		~ & ~ & ~ & 1 & 2.529 & 2.451 & 1.120  & 1.114  & 1.091  \\ 
		~ & ~ & ~ & 10 & 3.078 & 3.070 & 1.090  & 1.085  & 1.063  \\ 
		~ & ~ & 100 & 0.1 & 1.571 & 1.466 & 1.087  & 1.088  & 1.004  \\ 
		~ & ~ & ~ & 1 & 2.942 & 2.920 & 1.112  & 1.110  & 1.003  \\
		~ & ~ & ~ & 10 & 3.122 & 3.119 & 1.111  & 1.110  & 1.001  \\ 
		0.99 & 0.75 & 10 & 0.1 & 0.613 & 0.543 & 1.214  & 1.011  & 1.291  \\ 
		~ & ~ & ~ & 1 & 2.529 & 2.451 & 1.152  & 0.994  & 1.172  \\ 
		~ & ~ & ~ & 10 & 3.078 & 3.070 & 1.112  & 0.973  & 1.129  \\ 
		~ & ~ & 100 & 0.1 & 1.571 & 1.466 & 1.035  & 1.019  & 1.021  \\ 
		~ & ~ & ~ & 1 & 2.942 & 2.920 & 1.017  & 1.006  & 1.010  \\ 
		~ & ~ & ~ & 10 & 3.122 & 3.119 & 1.016  & 1.006  & 1.012  \\ 
		~ & 0.95 & 10 & 0.1 & 0.613 & 0.543 & 1.562  & 1.019  & 1.720  \\ 
		~ & ~ & ~ & 1 & 2.529 & 2.451 & 1.766  & 1.065  & 1.455  \\ 
		~ & ~ & ~ & 10 & 3.078 & 3.070 & 1.587  & 0.915  & 1.306  \\ 
		~ & ~ & 100 & 0.1 & 1.571 & 1.466 & 1.184  & 1.138  & 1.070  \\ 
		~ & ~ & ~ & 1 & 2.942 & 2.920 & 1.251  & 1.177  & 1.046  \\
		~ & ~ & ~ & 10 & 3.122 & 3.119 & 1.239  & 1.160  & 1.038  \\
		\hline
	\end{tabular}
\begin{tablenotes}
	\footnotesize              
	\item[1] $\phi^{I}$ is the angle $\phi$ in \eqref{cosine phi1} for Type $I$ bands ($SB$, $TBU$ and $TBE$ bands).          
	\item[2] $\phi^{II}$ is the angle $\phi$ in \eqref{cosine phi1} for Type $II$ bands ($V$, $UV$ and $TT$ bands). 
\end{tablenotes}
	\end{threeparttable}
\end{table}
\newpage

\newpage
\section{Real Data Example}
	
	In order to illustrate a visual demonstration of SCBs and their corresponding confidence sets $C(\bm{T}^*)$, a drug stability study based on the observations on the first batch of Experiment One in Ruberg and Hsu (1992) is used as a real data example.
	Drug stability studies are routinely carried out to understand the degradation over time of a drug product in the pharmaceutical industry. 
	One frequently used statistical model for drug stability studies is the simple linear regression. 
	According to the observations  in Ruberg and Hsu (1992), the fitted model is $\hat{y}=98.244-1.515(x-\bar{x})$ with the coefficient of determination $R^2=0.9961$, where $\bar{x}=\sum_{i=1}^n x_i/n=1.482$ and $n=9$.
	
	From patients' point of view, a large proportion, $1-\gamma$, of all the dosage units should have a drug content level above a pre-specified threshold $h$ before the expiry date. 
	In this case, it is desirable to establish the SCBs for $\bm{x}^T\bm{\beta}+z_{\gamma}\sigma$ to study the degradation of this drug product. 
	If the lower confidence band of $\bm{x}^T\bm{\beta}+z_{\gamma}\sigma$ is above the threshold $h$ (e.g., for a hypothetic value $h = 98$) over time $x\in(a,b)$, it signifies that $1-\gamma$ of all the dosage units have a drug content level above $h$ with confidence $1-\alpha$, and so this drug is still safe to use. 
	Since the expiry date required by the United States Food and Drug Administration (FDA) for drug package labels is generally no longer than two years, it is reasonable to set the time interval as $x\in (0, 2)$. 
	 
	Given the inputs $\bm{Y}$, $\bm{X}$, $(a, b)= (0-\bar{x}, 2-\bar{x})=(-1.482,0.518)$, $\gamma=0.05$ and $1-\alpha=0.95$, Table \ref{drug stability study} shows the ratios of areas of confidence sets among: (i) the symmetric SCBs and (ii) the asymmetric SCBs, over the covariate interval $(a, b)$. 
	It is clear from Table \ref{drug stability study} that the areas of confidence set for asymmetric SCBs are substantially smaller than those for symmetric SCBs under MACS criterion. 
	Based on the numerical results, we can conclude that UVa band is the best under MACS criterion in this example. 
	%Our computation for other data sets shows that the asymmetric SCBs can be narrower than the corresponding exact symmetric bands by more than the numbers given in Table \ref{drug stability study}. 
	
	\begin{table}[h]
		\centering
		\caption{\label{drug stability study}Ratios $r$ of the areas of confidence sets for the example on the drug stability study, relative to UVa band. }
				\begin{tabular}{cccccccc}
					\hline
					{Bands}&{Ratio}&{Bands}&{Ratio}&{Bands}&{Ratio}&{Bands}&{Ratio}\\
					\hline
					SB/UVa & 1.579  & V/UVa & 1.425  & SBa/UVa & 1.089  & Va/UVa & 1.002  \\ 
					TBU/UVa & 1.472  & UV/UVa & 1.335  & TBUa/UVa & 1.089  & UVa/UVa & 1.000  \\ 
					TBE/UVa & 1.235  & TT/UVa & 1.334  & TBEa/UVa & 1.091  & TTa/UVa & 1.000  \\ 
					\hline
			\end{tabular}
	\end{table}

\begin{figure}[h]
	\centering
	\includegraphics[width=10cm,height=9cm]{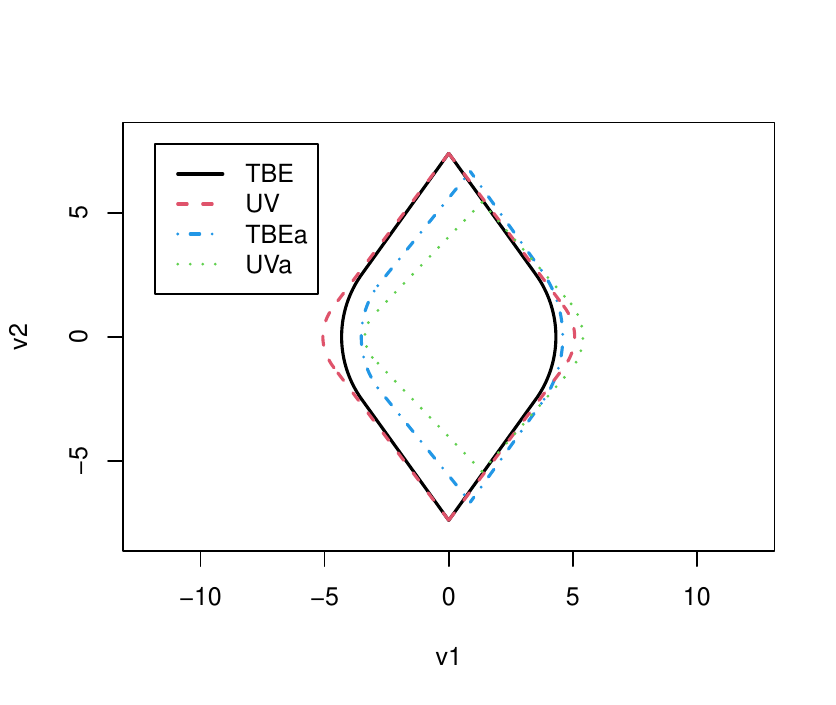}
	\caption{The areas of $C(\bm{T}^*)$ for bands TBE, UV, TBEa and UVa on coordinate system ($v_1$, $v_2$) }
	\label{drug eg figure scb}
\end{figure}

In order to visually demonstrate the disparity in the areas of confidence sets, $C(\bm{T}^*)$ for TBE, UV, TBEa and UVa are used in Figure \ref{drug eg figure scb}. 
The areas of $C(\bm{T}^*)$ for TBE, UV, TBEa and UVa are given by solid line, dash line, dash-doted line, and doted line, respectively. 
It is clear that the area of $C(\bm{T}^*)$ for UVa is the smallest.  
%Similar results are shown in Table  \ref{drug stability study}.

\begin{figure}[H]
	\centering
	\includegraphics[width=12cm,height=8cm]{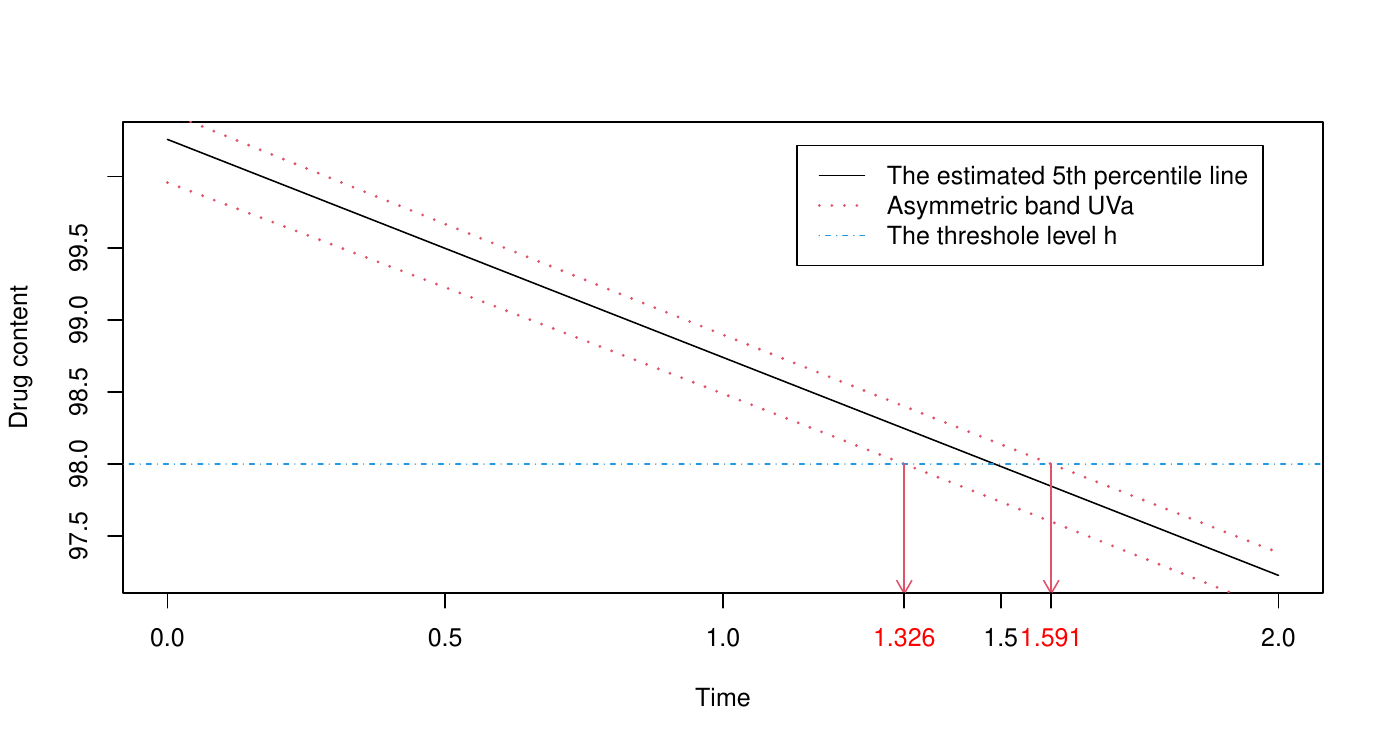}
	\caption{The 95\% asymmetric band UVa for the 5th percentile.}
	\label{drug eg figure}
\end{figure}

%In order to illustrate the drug stability study based on SCBs for percentiles in linear regression, UVa bands is used, which have been identified as the optimal choice in Table \ref{drug stability study}. 
In Figure \ref{drug eg figure}, the estimated percentile line $\bm{x}^T\hat{\bm{\beta}}+z_{\gamma}\hat{\sigma}/\theta$ is shown by the solid line, and the asymmetric Type $II$ band UVa is given by the doted lines.
The critical constants ($c_1$, $c_2$) for UVa are (3.230, 2.016). 
For the given threshold $h = 98$, which is given by the dash-dotted line, one can infer from the band UVa that, the percentile line $\bm{x}^T\bm{\beta}+z_{\gamma}\sigma$ is above $h$ before the time point $x = 1.326$, and so at least $1-\gamma=95\%$ proportion of all the dosage units have drug content above $h$ by this time point. 
But beyond the time point $x = 1.591$, the percentile line $\bm{x}^T\bm{\beta}+z_{\gamma}\sigma$ is below $h$, and so less than 95\% proportion of all the dosage units have drug content above $h$. 
It should be noted that the exact time point $x$ at which $\bm{x}^T\bm{\beta}+z_{\gamma}\sigma=h$ can be anywhere in the interval (1.326, 1.591). 

In general, according to the area of confidence sets in Figure \ref{drug eg figure scb} and the numerical results in Table \ref{drug stability study}, the asymmetric Type $II$ bands (Va, UVa and TTa) should be used under MACS criterion.  

\section{Conclusion}
The MACS criterion is proposed in this paper and it allows the comparison of different types of SCBs for percentiles based on the area of confidence set for the pivotal quantities $\bm{T}^*$ related to the unknown parameters $(\bm{\beta},\sigma)$. 
In addition, our newly proposed MACS-based method of calculating the critical constants for SCBs for percentiles is more computationally efficient than the previous methods available in the statistical literature. 
	
%For the comparison within each type of SCBs for percentiles, the results we get are consistent with the results based on the average width criterion in Han et al. (2015). 
In our paper, it is observed that the areas of confidence set for asymmetric SCBs are uniformly and can be very substantially smaller than those for the corresponding symmetric bands. 
%Asymmetric bands are better than the corresponding symmetric bands by construction under the MACS criterion, so asymmetric bands should always be chosen. 
Asymmetric bands are inherently superior to corresponding symmetric bands according to the MACS criterion, making them the preferred choice.
In most cases, asymmetric Type $II$ bands perform better. 
Therefore, asymmetric Type $II$ bands, like UVa, are recommended. 
	
Although this paper only focuses the simple linear regression model, the proposed method can readily be generalised to multiple linear regression and polynomial regression. 
%Also, it is possible to consider different covariate regions for multiple cases.  
Furthermore, the MACS criterion can be used to select optimal  simultaneous tolerance intervals for linear regression, which is currently under research and will be reported separately. 

%\noindent {\bf{Conflict of Interest}}
%
%\noindent {\it{The authors have declared no conflict of interest.}}

\begin{appendices}
	\section*{Appendix: Joint density function of $(V_1,V2)$ in $\bm{V}$}

Since $N_1\sim N(0,1)$, $N_2\sim N(0,1)$ and $U\sim \sqrt{\chi_{\nu}^2/\nu}$ are independent in \eqref{mixed asymmetric}, the joint density function of $(N_1,N_2,U)$ is 
\begin{equation}
	f(n_1,n_2,u)=\frac{\nu^{\nu/2}}{2^{\nu/2}\pi\Gamma(\nu/2)}u^{\nu-1}\exp(-\frac{1}{2}(n_1^2+n_2^2+\nu u^2)).
	\label{jdf for N and U}
\end{equation}
Now, consider the pdf for $\bm{V}$. We have 
\begin{equation*}
	V_1=q_1\frac{N_1-q_3}{U}+q_2\ \text{and}\ V_2=N_2/U,
\end{equation*}
where $q_1=\sqrt{\frac{1}{1+nz_{\gamma}^2\xi}}$,  $q_2=\frac{z_{\gamma}\sqrt{n}}{\sqrt{\theta^2(1+nz_{\gamma}^2\xi)}}$ and $q_3=z_{\gamma}\sqrt{n}$. 
Define the random variable $V_3=U$, such that the inverse functions $N_1$, $N_2$ and $U$ in terms of $V_1$, $V_2$ and $V_3$ are
\begin{equation}
	N_1=V_3\frac{V_1-q_2}{q_1}+q_3,\ N_2=V_2V_3,\ \text{and}\ U=V_3.
	\label{inverse function}
\end{equation}
This implies the following Jacobian matrix and determinant
\begin{align}
	\bm{J}&=\begin{pmatrix} \frac{\partial N_1}{\partial V_1} & \frac{\partial N_1}{\partial V_2} & \frac{\partial N_1}{\partial V_3}\\ \frac{\partial N_2}{\partial V_1} & \frac{\partial N_3}{\partial V_2} & \frac{\partial N_3}{\partial V_3}\\ \frac{\partial U}{\partial V_1} & \frac{\partial U}{\partial V_2} & \frac{\partial U}{\partial V_3} \end{pmatrix}=\begin{pmatrix} \frac{V_3}{q_1} & 0 & \frac{V_1-q_2}{q_1} \\ 0 & V_3 & V_2 \\ 0 & 0 & 1 \end{pmatrix},\nonumber\\ |\bm{J}|&=V_3^2/q_1.
	\label{jacobian determinant}
\end{align}
With the probability density function of an invertible function, the joint density of $V_1$, $V_2$ and $V_3$ can be derived as 
\begin{equation*}
	f_{V_1,V_2,V_3}(v_1,v_2,v_3)=f_{\bm{N},U}(n_1,n_2,u)|J|.
\end{equation*}
Substituting \eqref{inverse function} into \eqref{jdf for N and U}, and then with \eqref{jacobian determinant}, we have
\begin{align*}
	f_{V_1,V_2,V_3}(v_1,v_2,v_3)=\frac{\nu^{\nu/2}}{q_1 2^{\nu/2}\pi\Gamma(\nu/2)}v_3^{\nu+1}\exp\Bigg\{-\frac{1}{2}\Bigg[&\left(\left(\frac{v_1-q_2}{q_1}\right)^2+v_2^2+\nu\right)v_3^2\nonumber\\
	&+\frac{2q_3(v_1-q_2)}{q_1}v_3+q_3^2\Bigg]\Bigg\}.
\end{align*}
The marginal density of $V_1$ and $V_2$ can be obtained by integrating out $V_3$
\begin{align*}
	f_{V_1,V_2}(v_1,v_2)=\frac{\exp(-q_3^2/2)\nu^{\nu/2}}{q_1 2^{\nu/2}\pi\Gamma(\nu/2)}\int_{0}^{\infty}v_3^{\nu+1}\exp\Bigg\{-\frac{1}{2}\Bigg[&\left(\left(\frac{v_1-q_2}{q_1}\right)^2+v_2^2+\nu\right)v_3^2\nonumber\\
	&+\frac{2q_3(v_1-q_2)}{q_1}v_3\Bigg]\Bigg\}dv_3.
\end{align*}

\end{appendices}
	
\vskip 12pt
\noindent{\bf  Reference}
%\vskip 7pt

%\vskip 2pt
%\noindent
%Ah-Kine, P., and Liu, W. (2015). Optimal simultaneous confidence bands in multiple linear regression with predictor variables constrained in an ellipsoidal region. {\it Communications in Statistics-Theory and Methods}, 44(3), 441-452.

\vskip 2pt
\noindent
Al-Saidy, O. M., Piegorsch, W. W., West, R. W., and Nitcheva, D. K. (2003). Confidence bands for low‐dose risk estimation with quantal response data. {\it Biometrics}, 59(4), 1056-1062.

%\vskip 2pt
%\noindent
%Atkinson, A., Donev, A., and Tobias, R. (2007). {\it Optimum experimental designs, with SAS} (Vol. 34). OUP Oxford.

\vskip 2pt
\noindent
Gafarian, A. V. (1964). Confidence bands in straight line regression. {\it Journal of the American Statistical Association}, 59(305), 182-213.

\vskip 2pt
\noindent
Han, Y., Liu, W., Bretz, F., and Wan, F. (2015). Simultaneous confidence bands for a percentile line in linear regression. {\it Computational Statistics and Data Analysis}, 81, 1-9.

\vskip 2pt
\noindent
Liu, W., and Ah-kine, P. (2010). Optimal simultaneous confidence bands in simple linear regression. {\it Journal of statistical planning and inference}, 140(5), 1225-1235.

\vskip 2pt
\noindent
Liu, W., Bretz, F., Hayter, A. J., and Wynn, H. P. (2009). Assessing nonsuperiority, noninferiority, or equivalence when comparing two regression models over a restricted covariate region. {\it Biometrics}, 65(4), 1279-1287.

\vskip 2pt
\noindent
Liu, W., and Hayter, A. J. (2007). Minimum area confidence set optimality for confidence bands in simple linear regression. {\it Journal of the American Statistical Association}, 102(477), 181-190.

\vskip 2pt
\noindent
Liu, W., Jamshidian, M., and Zhang, Y. (2004). Multiple comparison of several linear regression models. {\it Journal of the American Statistical Association}, 99(466), 395-403.

\vskip 2pt
\noindent
Liu, W., Jamshidian, M., Zhang, Y., Bretz, F., and Han, X. L. (2007). Pooling batches in drug stability study by using constant‐width simultaneous confidence bands. {\it Statistics in medicine}, 26(14), 2759-2771.

\vskip 2pt
\noindent
Liu, W., Lin, S., and Piegorsch, W. W. (2008). Construction of exact simultaneous confidence bands for a simple linear regression model. {\it International Statistical Review}, 76(1), 39-57.

\vskip 2pt
\noindent
Piegorsch, W. W., Webster West, R., Pan, W., and Kodell, R. L. (2005). Low dose risk estimation via simultaneous statistical inferences. {\it Journal of the Royal Statistical Society: Series C (Applied Statistics)}, 54(1), 245-258.

\vskip 2pt
\noindent
Ruberg, S. J., and Hsu, J. C. (1992). Multiple comparison procedures for pooling batches in stability studies. {\it Technometrics}, 34(4), 465-472.

\vskip 2pt
\noindent
Spurrier, J. D. (1999). Exact confidence bounds for all contrasts of three or more regression lines. {\it Journal of the American Statistical Association}, 94(446), 483-488.

\vskip 2pt
\noindent
Steinhorst, R. K., and Bowden, D. C. (1971). Discrimination and confidence bands on percentiles. {\it Journal of the American Statistical Association}, 66(336), 851-854.

\vskip 2pt
\noindent
Thomas, D. L., and Thomas, D. R. (1986). Confidence bands for percentiles in the linear regression model. {\it Journal of the American Statistical Association}, 81(395), 705-708.

\vskip 2pt
\noindent
Turner, D. L., and Bowden, D. C. (1977). Simultaneous confidence bands for percentile lines in the general linear model. {\it Journal of the American Statistical Association}, 72(360a), 886-889.

\vskip 2pt
\noindent
Turner, D. L., and Bowden, D. C. (1979). Sharp confidence bands for percentile lines and tolerance bands for the simple linear model. {\it Journal of the American Statistical Association}, 74(368), 885-888.

\end{document}